%% file: ulc_variableB.tex
\documentclass[conference]{IEEEtran}
\IEEEoverridecommandlockouts

\usepackage{amsmath,amsfonts}
\usepackage{algorithmic}
\usepackage{array}
\usepackage[caption=false,font=normalsize,labelfont=sf,textfont=sf]{subfig}
\usepackage{textcomp}
\usepackage{stfloats}
\usepackage{url}
\usepackage{verbatim}
\usepackage{graphicx}
\def\BibTeX{{\rm B\kern-.05em{\sc i\kern-.025em b}\kern-.08em
    T\kern-.1667em\lower.7ex\hbox{E}\kern-.125emX}}
\usepackage{balance}
\usepackage{caption}
\usepackage{siunitx}
\usepackage{tikz}
\usetikzlibrary{shapes.geometric, arrows.meta, positioning, calc, fit}

\usepackage{algorithm}
\input{miko}

\usetikzlibrary{intersections,pgfplots.fillbetween,patterns,spy}

\Crefname{figure}{Fig.}{Figs.}
\pgfplotsset{compat=1.15}

\ifdefined\ONECOLUMN
\newcommand{\smallplotheight}{0.\columnwidth} 
\newcommand{\smallplotwidth}{0.375\columnwidth}
\newcommand{\normalplotheight}{0.25\columnwidth}
\newcommand{\normalplotwidth}{0.5\columnwidth}
\newcommand{\largeplotheight}{0.36\columnwidth}
\else
\newcommand{\smallplotheight}{0.45\columnwidth} 
\newcommand{\smallplotwidth}{0.725\columnwidth}
\newcommand{\normalplotheight}{0.45\columnwidth}
\newcommand{\normalplotwidth}{1.0\columnwidth}
\newcommand{\largeplotheight}{0.75\columnwidth}
\newcommand{\largeplotwidth}{1.0\columnwidth}
\fi

\newcommand{\lineWidth}{1.0pt}
\newcommand{\markSize}{2.0pt}
\definecolor{ourdarkblue}{RGB}{30, 100, 200}
\definecolor{ourdarkgreen}{RGB}{0, 100, 0}
\definecolor{ourdarkorange}{RGB}{201, 98, 18}
\definecolor{ouryellow}{RGB}{220, 210, 50}

\tikzset{unipowcdf/.style={mark options={solid}, color=ourdarkgreen, line width=\lineWidth, mark=None, mark size=\markSize, dotted}}
\tikzset{upchansubcdf/.style={mark options={solid}, color=purple, line width=\lineWidth, mark=None, mark size=\markSize, dotted}}

\tikzset{lloydpgddl/.style={mark options={solid}, color=TUMBeamerOrange, line width=\lineWidth, mark=None, mark size=\markSize}}
\tikzset{lloydpgdul/.style={mark options={solid}, color=cyan, line width=\lineWidth, mark=None, mark size=\markSize, dotted}}
\tikzset{lloydlaudl/.style={mark options={solid}, color=TUMBeamerRed, line width=\lineWidth, mark=None, mark size=\markSize}}
\tikzset{lloydlauul/.style={mark options={solid}, color=TUMBeamerGreen, line width=\lineWidth, mark=None, mark size=\markSize, dotted}}
\tikzset{gmmpgddl/.style={mark options={solid}, color=TUMBlue, line width=\lineWidth, mark=None, mark size=\markSize}}
\tikzset{gmmpgdul/.style={mark options={solid}, color=brown, line width=\lineWidth, mark=None, mark size=\markSize, dotted}}
\tikzset{gmmlaudl/.style={mark options={solid}, color=gray, line width=\lineWidth, mark=None, mark size=\markSize}}
\tikzset{gmmlauul/.style={mark options={solid}, color=black, line width=\lineWidth, mark=None, mark size=\markSize, dotted}}

\tikzset{gmmmrgmrg/.style={mark options={solid}, color=gray, line width=\lineWidth, mark=None, mark size=\markSize}}

\newcommand{\pltHeigsp}{directions}

\tikzset{lloydpgdulperfect/.style={mark options={solid}, color=gray, line width=\lineWidth, mark=None, mark size=\markSize, dashed}}
\tikzset{lloydpgdulomp/.style={mark options={solid}, color=TUMBeamerOrange, line width=\lineWidth, mark=None, mark size=\markSize, dashed}}
\tikzset{lloydpgdulls/.style={mark options={solid}, color=red, line width=\lineWidth, mark=None, mark size=\markSize}}
\tikzset{lloydpgdulscov/.style={mark options={solid}, color=TUMBeamerRed, line width=\lineWidth, mark=None, mark size=\markSize, dashed}}
\tikzset{lloydpgdulgmmulbest/.style={mark options={solid}, color=gray, line width=\lineWidth, mark=None, mark size=\markSize}}
\tikzset{lloydpgdulgmmulall/.style={mark options={solid}, color=black, line width=\lineWidth, mark=None, mark size=\markSize, dashed}}

\tikzset{gmmpgdulperfect/.style={mark options={solid}, color=TUMBeamerBlue, line width=\lineWidth, mark=None, mark size=\markSize}}
\tikzset{gmmpgdulfromy/.style={mark options={solid}, color=TUMBeamerGreen, line width=\lineWidth, mark=None, mark size=\markSize}}

\newcommand{\pltlloydpgdulomp}{Lloyd, $\hhat_{\text{OMP}}$}
\newcommand{\pltlloydpgdulscov}{Lloyd, $\hhat_{\text{LMMSE}}$}
\newcommand{\pltlloydpgdulgmmulall}{Lloyd, $\hhat_{\text{GMM}}$}

\newcommand{\pltgmmpgdulfromy}{GMM, $\mby$}

\tikzset{MUlloydpgdulperfect/.style={mark options={solid}, color=gray, line width=\lineWidth, mark=None, mark size=\markSize, dashed}}
\tikzset{MUlloydpgdulomp/.style={mark options={solid}, color=TUMBeamerOrange, line width=\lineWidth, mark=None, mark size=\markSize, dashed}}
\tikzset{MUlloydpgdulscov/.style={mark options={solid}, color=TUMBeamerRed, line width=\lineWidth, mark=None, mark size=\markSize, dashed}}
\tikzset{MUlloydpgdulgmmulall/.style={mark options={solid}, color=black, line width=\lineWidth, mark=None, mark size=\markSize, dashed}}

\tikzset{MUrandomperfect/.style={mark options={solid}, color=black, line width=\lineWidth, mark=None, mark size=\markSize, dotted}}
\tikzset{MUrandomomp/.style={mark options={solid}, color=TUMBeamerOrange, line width=\lineWidth, mark=None, mark size=\markSize, dotted}}
\tikzset{MUrandomscov/.style={mark options={solid}, color=TUMBeamerRed, line width=\lineWidth, mark=None, mark size=\markSize, dotted}}
\tikzset{MUrandomgmmulall/.style={mark options={solid}, color=gray, line width=\lineWidth, mark=None, mark size=\markSize, dotted}}

\tikzset{MUgmmpgdulperfect/.style={mark options={solid}, color=TUMBeamerBlue, line width=\lineWidth, mark=None, mark size=\markSize}}
\tikzset{MUgmmpgdulfromy/.style={mark options={solid}, color=TUMBeamerGreen, line width=\lineWidth, mark=None, mark size=\markSize}}

\tikzset{MUswmmsegmmpgdulperfect/.style={mark options={solid}, color=TUMBeamerBlue, line width=\lineWidth, mark=None, mark size=\markSize, dashdotted}}
\tikzset{MUswmmsegmmpgdulfromy/.style={mark options={solid}, color=TUMBeamerGreen, line width=\lineWidth, mark=None, mark size=\markSize, dashdotted}}

\newcommand{\pltMUlloydpgdulomp}{Lloyd, $\hhat_{\text{OMP}}$}
\newcommand{\pltMUlloydpgdulscov}{Lloyd, $\hhat_{\text{LMMSE}}$}
\newcommand{\pltMUlloydpgdulgmmulall}{Lloyd, $\hhat_{\text{GMM}}$}

\newcommand{\pltMUrandomgmmulall}{Random, $\hhat_{\text{GMM}}$}

\newacronym{AWGN}{AWGN}{additive white Gaussian noise}
\newacronym{BLMMSE}{BLMMSE}{Bussgang LMMSE}
\newacronym{BS}{BS}{base station}
\newacronym{CDF}{CDF}{cumulative distribution function}
\newacronym{CNN}{CNN}{convolutional neural network}
\newacronym{CSI}{CSI}{channel state information}
\newacronym{CSIT}{CSIT}{channel state information at the transmitter}
\newacronym{DFT}{DFT}{discrete Fourier transform}
\newacronym{DL}{DL}{downlink}
\newacronym{DNN}{DNN}{deep neural network}
\newacronym{DoA}{DoA}{direction of arrival}
\newacronym{EM}{EM}{expectation maximization}
\newacronym{FDD}{FDD}{frequency division duplex}
\newacronym{GMM}{GMM}{Gaussian mixture model}
\newacronym{LMMSE}{LMMSE}{linear minimum mean square error}
\newacronym{LOS}{LOS}{line of sight}
\newacronym{LS}{LS}{least squares}
\newacronym{MSE}{MSE}{mean squared error}
\newacronym{MIMO}{MIMO}{multiple-input multiple-output}
\newacronym{MPC}{MPC}{multi-path component}
\newacronym{MT}{MT}{mobile terminal}
\newacronym{NLOS}{NLOS}{non-line of sight}
\newacronym{NN}{NN}{neural network}
\newacronym{O2I}{O2I}{outdoor-to-indoor}
\newacronym{OMP}{OMP}{orthogonal matching pursuit}
\newacronym{PDF}{PDF}{probability density function}
\newacronym{PGA}{PGA}{projected gradient ascent}
\newacronym{PSD}{PSD}{power spectral density}
\newacronym{SNR}{SNR}{signal-to-noise ratio}
\newacronym{TDD}{TDD}{time division duplex}
\newacronym{UL}{UL}{uplink}
\newacronym{ULA}{ULA}{uniform linear array}
\newacronym{URA}{URA}{uniform rectangular array}
\newacronym{UMa}{UMa}{urban macrocell}
\newacronym{nSE}{nSE}{normalized spectral efficiency}
\newacronym{cCDF}{cCDF}{complementary cumulative distribution function}
\newacronym{MU-MIMO}{MU-MIMO}{multi-user MIMO}
\newacronym{MU-MISO}{MU-MISO}{multi-user MISO}
\newacronym{BD}{BD}{block diagonalization}
\newacronym{RBD}{RBD}{regularized block diagonalization}
\newacronym{RCI}{RCI}{regularized channel inversion}
\newacronym{WMMSE}{WMMSE}{weighted minimum mean square error}
\newacronym{SWMMSE}{SWMMSE}{stochastic WMMSE}
\newacronym{SVD}{SVD}{singular value decomposition}
\newacronym{SR}{SR}{sum-rate}
\newacronym{CME}{CME}{conditional mean estimator}
\newacronym{ML}{ML}{machine learning}
\newacronym{FLOPS}{FLOPS}{floating-point operations}

\newcommand{\Nrx}{N_{\mathrm{rx}}}
\newcommand{\Ntx}{N_{\mathrm{tx}}}
\newcommand{\Ntxv}{N_{\mathrm{tx,v}}}
\newcommand{\Ntxh}{N_{\mathrm{tx,h}}}
\newcommand{\Krx}{K_{\mathrm{rx}}}
\newcommand{\Ktx}{K_{\mathrm{tx}}}

\begin{document}


\title{Enhanced Low-Complexity FDD System Feedback with Variable Bit Lengths via Generative Modeling}
\author{Nurettin~Turan, Benedikt~Fesl, and Wolfgang~Utschick\\
\IEEEauthorblockA{TUM School of Computation, Information and Technology, Technical University of Munich, Germany\\
	Email: \{nurettin.turan, benedikt.fesl, utschick\}@tum.de
    }
\thanks{The authors acknowledge the financial support by the Federal Ministry of
Education and Research of Germany in the program of ``Souver\"an. Digital.
Vernetzt.''. Joint project 6G-life, project identification number: 16KISK002.}
\thanks{\copyright This work has been submitted to the IEEE for possible publication. Copyright may be transferred without notice, after which this version may no longer be accessible.
}
}


\maketitle

\begin{abstract}
    Recently, a versatile limited feedback scheme based on a \ac{GMM} was proposed for \ac{FDD} systems.
    This scheme provides high flexibility regarding various system parameters and is applicable to both point-to-point \ac{MIMO} and \ac{MU-MIMO} communications. 
    The \ac{GMM} is learned to cover the operation of all \acp{MT} located inside the \ac{BS} cell, and each \ac{MT} only needs to evaluate its strongest mixture component as feedback, eliminating the need for channel estimation at the MT.
    In this work, we extend the \ac{GMM}-based feedback scheme to variable feedback lengths by leveraging a single learned \ac{GMM} through merging or pruning of dispensable mixture components. 
    Additionally, the \ac{GMM} covariances are restricted to Toeplitz or circulant structure through model-based insights. 
    These extensions significantly reduce the offloading amount and enhance the clustering ability of the \ac{GMM} which, in turn, leads to an improved system performance. 
    Simulation results for both point-to-point and multi-user systems demonstrate the effectiveness of the proposed extensions.
\end{abstract}


\section{Introduction}
In the next generation of cellular systems (6G), the \ac{BS} has the ability to adjust to changing channel conditions. 
However, in \ac{FDD} systems, this adaptation must rely on feedback from the \ac{MT} since channel reciprocity is not maintained~\cite{Love}. 
There is considerable interest in systems that use limited feedback, where only a few bits are designated \cite{Love}.
In this context, two main strategies can be distinguished.
The first involves estimating the \ac{DL} channel at the \acp{MT} and determining the feedback based on this \cite{Love, RaJi08, LaYoCh04}. The second approach aims to directly encode feedback from pilot observations, such as through deep learning, cf., e.g., \cite{JaLeKiLe22, TuKoBaXuUt21}.

In recent works, a versatile \ac{GMM}-based limited feedback scheme was proposed which provides flexibility with respect to the number of antennas, the transmission mode, the number of \acp{MT}, the supported \ac{SNR} range, the number of pilots, and the choice of the precoding algorithm, together with low complexity to determine the feedback by circumventing the necessity for channel estimation at the \acp{MT} \cite{TuKoFeBaXuUt22,TuFeKoJoUt23}.
The \acp{MT}, select the index of the \ac{GMM} component with the highest responsibility (posterior probability) for their received pilot signal as their feedback information.
The corresponding \ac{GMM} is learned from data which represent the underlying channel distribution of a whole communication scenario inside a \ac{BS} cell. 
This is motivated by the universal approximation property \cite{NgNgChMc20} and the strong results of \acp{GMM} in wireless communications \cite{KoFeTuUt21J,10078293,FeJoHuKoTuUt22,TuFeGrKoUt22}. 
A main advantage is that the offline fitting of the \ac{GMM} can be done centralized at the \ac{BS} due to the absence of a distributional shift between the channel distributions of \ac{UL} and \ac{DL}, cf.~\cite{utschick2021,TuKoRiFeBaXuUt21,fesl2021centralized}.

Although \acp{GMM} possess the universal approximation property, cf. \cite{NgNgChMc20}, a known trait is that their corresponding mixture components may not be distinct enough from each other to be interpreted as clusters \cite{He10}. 
This similarly also arises in the \ac{GMM}-based feedback scheme from \cite{TuFeKoJoUt23}, where the number $K$ of mixture components is predetermined by the number $B$ of feedback bits, i.e., $K=2^B$. Therefore, the \ac{GMM} should exhibit both a strong clustering ability for selecting the feedback index and a good likelihood model for the underlying channel distribution of the communication scenario, while maintaining a fixed number of components. 
To address this problem, various merging and pruning techniques have been proposed in the literature \cite{Ru07,He10}.
Furthermore, structural features of the covariance matrices, imposed by the antenna arrays, can be utilized in order to reduce the number of parameters of the \ac{GMM} which, in turn, leads to a higher robustness against overfitting for a limited amount of training data \cite{FeJoHuKoTuUt22}.

\emph{Contributions:} 
This work extends the versatile \ac{GMM}-based feedback scheme from \cite{TuKoFeBaXuUt22,TuFeKoJoUt23} by leveraging a single learned \ac{GMM} to variable feedback bit lengths through merging or pruning of dispensable mixture components. 
Thereby, a simple pruning technique and a merging approach---which combines components with high similarity---is analyzed.
Additionally, we investigate the restriction of the \ac{GMM} to Toeplitz- and circulant-structured covariances which drastically reduces the necessary offloading overhead from the \ac{BS} to the \acp{MT} and prevents overfitting in the case of limited training data.
Simulation results show that the reduction of \ac{GMM} components through merging or pruning enables variable bit lengths and is superior to directly fitting a smaller \ac{GMM} because of the enhanced clustering ability. 
Furthermore, the structured \ac{GMM} variants yield great performances with low memory overhead for both point-to-point and \ac{MU-MIMO} systems.

\section{System and Channel Models}
\label{sec:system_channel_model}

\subsection{Data Transmission Phase -- Point-to-Point MIMO System}

In a point-to-point \ac{MIMO} system the \ac{DL} received signal is $\mby^\prime = \mbH \mbx + \mbn^\prime$, where $\mby^\prime \in \C^{\Nrx}$ is the receive vector, \( \mbx = \mbQ^{1/2} \mbs \in \C^{\Ntx}\) with \( \expec[\mbs \mbs^\herm] = \mathbf{I}_{\Ntx} \) is the precoded transmit vector, $\mbH \in \C^{\Nrx \times \Ntx}$ is the \ac{MIMO} channel with $\Nrx < \Ntx$, and $\mbn^\prime \sim \mathcal{N}_\C(\mathbf{0},
\sigma_n^2 \mathbf{I}_{\Nrx})$ denotes the \ac{AWGN}.
In a limited feedback system, \( B \) bits are used for encoding the feedback index $k^\star \in \{1,2, \cdots, 2^B\}$ that specifies an element from a codebook of $ 2^B $ pre-computed transmit covariance matrices $ \mathcal{Q} = \{\mbQ_1, \mbQ_2, \dots, \mbQ_{2^B} \} $ \cite{Love}.


\subsection{Data Transmission Phase -- Multi-user MIMO System}
\label{sec:conv_mumimo}

We utilize linear precoding in a single-cell \ac{MU-MIMO} \ac{DL} system.
The \ac{BS} is equipped with $\Ntx$ transmit antennas and each \ac{MT} $j \in \mathcal{J} = \{1, 2, \dots, J\}$, is equipped with $\Nrx$ antennas.
The precoded \ac{DL} data vector is $\mbx = \sum_{j=1}^{J}\mbM_j\mbs_j$, where $\mbs_j \in \C^{d_j}$ is the transmit signal of \ac{MT} $j$, with $\expec[\mbs_j]=\mathbf{0}$ and $\expec[\mbs_j\mbs_j^\herm]=\mathbf{I}_{d_j}$, and $\mbM_j \in \C^{\Ntx \times d_j}$ is the precoding matrix of \ac{MT} $j$.
The precoders fulfill $\operatorname{tr}(\sum_{j=1}^J\mbM_j \mbM_j^\herm) = \rho$.
In the multi-user setup, each \ac{MT} reports its own feedback information $k^\star_j$ to the \ac{BS}, which then jointly designs the precoders $\mbM_j$.


\subsection{Pilot Transmission Phase}


During the pilot transmission phase, every \ac{MT} indexed as $j \in \mathcal{J}$ receives
\begin{equation} \label{eq:noisy_obs}
    \mbY_j = \mbH_j \mbP + \mbN_j \in \C^{\Nrx \times n_p}
\end{equation}
where $\mbN_j = [\mbn^{\prime}_{j,1}, \dots, \mbn^{\prime}_{j,n_p}] \in \C^{\Nrx \times n_p}$ with $\mbn^\prime_{j,p} \sim \mathcal{N}_\C(\bm{0}, \sigma_j^2 \mathbf{I}_{\Nrx})$, for $p \in \{1,2, \dots, n_p\}$ and $n_p$ is the number of pilots, where we consider $n_p\leq \Ntx$.
It is assumed that the \ac{BS} employs a \ac{URA}.
Hence, we utilize a $2$D-DFT sub-matrix as the pilot matrix, cf., e.g., \cite{TuFeKoJoUt23}.
To maintain the power constraint each column $\mbp_p$ of $\mbP$ is normalized: $\|\mbp_p\|^2=\rho$.
For the subsequent analysis, it is advantageous to vectorize~\eqref{eq:noisy_obs}: $\mby_j = \mbA \mbh_j + \mbn_j$, where \( \mbh_j = \vect(\mbH_j) \), \( \mby_j = \vect(\mbY_j) \), \( \mbn_j = \vect(\mbN_j) \), \( \mbA = \mbP^\tp \otimes \mathbf{I}_{\Nrx} \) and $\mbn_j \sim \mathcal{N}_\C(\mathbf{0}, \mbSigma)$ with $\mbSigma = \sigma_n^2 \mathbf{I}_{\Nrx n_p}$.
In case of a point-to-point \ac{MIMO} system, we simplify notation for convenience by omitting the index $j$, resulting in
\begin{equation} \label{eq:noisy_obs_p2p}
    \mby = \mbA \mbh + \mbn \in \C^{\Nrx n_p}.
\end{equation}
To simplify notation further, we will use the channel matrix $\mbH$ and its vectorized form $\mbh$ interchangeably.

\subsection{Channel Model and Data Generation} \label{sec:data_generation}

The QuaDRiGa channel simulator \cite{QuaDRiGa1} is used to create a training dataset 
\begin{equation} \label{eq:H_dataset}
     \mathcal{H} = \{\mbh_\ell = \vect{(\mbH_\ell)}\}_{\ell=1}^{L}
\end{equation}
consisting of $L$ channel realizations for the above system model. 
Since we utilize the same simulation setup, we refer the reader to \cite{TuFeKoJoUt23} for more details.

\section{GMM-based Limited Feedback Scheme}


\label{sec:proposedscheme}

\subsection{GMM Preliminaries}

A \ac{GMM} is a \ac{PDF} of the form
\begin{equation}\label{eq:gmm_of_h}
    f^{(K)}_{\mbh}(\mbh) = \sum\nolimits_{k=1}^K \pi_k \calN_{\C}(\mbh; \mbmu_k, \mbC_k)
\end{equation}
where each term within the summation represents one of its \( K \) constituent \textit{components}.
The maximum likelihood estimates for the \ac{GMM}'s parameters, which include the mixing coefficients $\pi_k$, means $\mbmu_k$, and covariances $\mbC_k$, can be calculated by utilizing a training dataset \(\mathcal{H} \) as described in \eqref{eq:H_dataset}, along with an \ac{EM} algorithm outlined in \cite[Subsec.~9.2.2]{bookBi06}.
\acp{GMM} enable the computation of \textit{responsibilities}~\cite[Sec.~9.2]{bookBi06},
\begin{equation}\label{eq:responsibilities_h}
    p(k \mid \mbh) = \frac{\pi_k \calN_{\C}(\mbh; \mbmu_k, \mbC_k)}{\sum_{i=1}^K \pi_i \calN_{\C}(\mbh; \mbmu_i, \mbC_i) }
\end{equation}
which represent the posterior probability that a particular \( \mbh \) originates from component \( k \).
The inherent joint Gaussianity of each \ac{GMM} component and the \ac{AWGN} enables the simple computation of the \ac{GMM} of the observations by utilizing the \ac{GMM} from \eqref{eq:gmm_of_h} as
\begin{equation}\label{eq:gmm_y}
    f_{\mby}^{(K)}(\mby) = \sum\nolimits_{k=1}^K \pi_k \calN_{\C}(\mby; \mbA \meanhk, \mbA \covhk \mbA^\herm + \mbSigma).
\end{equation}
Analogously, we can compute:
\begin{equation}\label{eq:responsibilities}
    p(k \mid \mby) = \frac{\pi_k \calN_{\C}(\mby; \mbA \meanhk, \mbA \covhk \mbA^\herm + \mbSigma)}{\sum_{i=1}^K \pi_i \calN_{\C}(\mby; \mbA \meanhi, \mbA \covhi \mbA^\herm + \mbSigma) }.
\end{equation}

\subsection{Point-to-Point MIMO System}
\label{sec:coebookconstruction}

In an offline phase, firstly, a codebook \( \mc{Q} = \{ \mbQ_k \}_{k=1}^K \), with $K=2^B$, is constructed based on the \ac{GMM} \( f_{\mbh}^{(K)} \) from \eqref{eq:gmm_of_h} that is learned from the dataset \eqref{eq:H_dataset}.
Therefore, by utilizing the \ac{GMM}, the training data are clustered according to their \ac{GMM} responsibilities.
That is, \( \mc{H} \) is partitioned into \( K \) disjoint sets
\begin{equation}\label{eq:gmmcb_stage_1}
    \mc{V}_k = \{ \mbh \in \mathcal{H} \mid p(k \mid \mbh) \geq p(j \mid \mbh) \text{ for } k\neq j \}
\end{equation}
for \( k = 1, \dots, K \).
Then each codebook entry is determined:
\ifdefined\ONECOLUMN
\begin{equation}\label{eq:gmmcb_stage_2}
    \mbQ_k = \argmax_{\mbQ \succeq \mbzero} \frac{1}{|\mathcal{V}_k|} \sum_{\vect(\mbH)\in\mathcal{V}_k} r(\mbH,\mbQ) \quad \text{s.t.} \quad \operatorname{tr}(\mbQ) \leq \rho
\end{equation}
\else
\begin{align}\label{eq:gmmcb_stage_2}
    &\mbQ_k = \argmax_{\mbQ \succeq \mbzero} \tfrac{1}{|\mathcal{V}_k|} \sum_{\vect(\mbH)\in\mathcal{V}_k} r(\mbH,\mbQ) \\
    & \text{subject to} \quad \operatorname{tr}(\mbQ) \leq \rho \nonumber
\end{align}
\fi
where the spectral efficiency is
\begin{equation}
    r(\mbH, \mbQ) = \log_2 \det\left( \mathbf{I} + \tfrac{1}{\sigma_n^2} \mbH \mbQ \mbH^\herm\right).
    \label{speceff}
\end{equation}
A projected gradient ascent algorithm is utilized to solve this optimization problem, cf.~\cite{TuFeKoJoUt23}.
In the online phase, explicit channel estimation is circumvented and the pilot observation \( \mby \) is used to determine the feedback index via:
\begin{equation} \label{eq:ecsi_index}
    k^\star = \argmax_{k } {~p(k \mid \mby)}.
\end{equation}
The computational complexity for determining the feedback via \eqref{eq:ecsi_index} by using the \ac{GMM} is \( \calO(K \Nrx^2 n_p^2) \), cf. \cite{TuFeKoJoUt23}.
Thus, the complexity is not affected by $\Ntx$, which is particularly beneficial for massive \ac{MIMO} systems.
It's worth noting that parallelization is possible, given that the evaluation of all $K$ responsibilities can be conducted independently.
The \ac{MT} solely needs the \ac{GMM} parameters for calculating \eqref{eq:ecsi_index} and does not require any codebook knowledge.
Furthermore, the \ac{GMM} of the observations in \eqref{eq:gmm_y} can be easily customized at the \ac{MT} to accommodate various \ac{SNR} levels and pilot configurations, by simply updating the means and covariances, cf. \eqref{eq:gmm_y}, without the need for retraining.
As a baseline for performance analysis, perfect \ac{CSI} can be used to determine the feedback
\begin{equation} \label{eq:pcsi_index}
    k^\star = \argmax_{k } ~{p(k \mid \mbh)}.
\end{equation}


\subsection{Multi-User MIMO System}
\label{sec:mumimo}

In \cite{TuFeKoJoUt23} it was shown that directional information associated with each codebook entry can be obtained by conducting a \ac{SVD}, i.e., $\mbQ_k = \mbX_{k} \mbT_k \mbX_{k}^\herm$.
Thereby, the matrix $\mbT_k$ arranges the singular values in descending order, while the matrix $\bar{\mbX}_k$ encompasses the first $\Nrx$ vectors of $\mbX_{k}$ as the relevant directional information.
Consequently, the set $\mc{Q} = \{\bar{\mbX}_1, \bar{\mbX}_2, \cdots, \bar{\mbX}_{K}\}$ forms a directional codebook, cf.~\cite{TuFeKoJoUt23}.
Utilizing the \ac{GMM}-based approach each \ac{MT} efficiently computes its feedback by:
\begin{equation} \label{eq:ecsi_index_j}
    k^\star_j = \argmax_{k } ~{p(k \mid \mby_j)}.
\end{equation}
Every \ac{MT} transmits the index $k^\star_j$ to the \ac{BS}, which subsequently represents each \ac{MT}'s channel with the subspace information linked with the corresponding codebook entry $\widetilde{\mbH}_j = \bar{\mbX}^\herm_{k^\star_j}$.
The \ac{BS} can then utilize widely-used precoding techniques, such as the iterative \ac{WMMSE}~\cite{HuCaYuQiXuYuDi21}, in order to jointly design the precoders, cf.~\cite{TuFeKoJoUt23}.

Alternatively, an approach based on generative modeling was also introduced in \cite{TuFeKoJoUt23}.
In this approach, the channel matrix of each \ac{MT} is treated as a random variable, and the precoders are computed using the \ac{SWMMSE} algorithm~\cite{RaBoLu13}.
The \ac{GMM}-based approach is able to generate samples that adhere to the channel's distribution due to the inherent sample generation capability of \acp{GMM}.
Specifically, given the feedback $k_j^\star$ of each \ac{MT}, see \eqref{eq:ecsi_index_j}, one can generate samples through $\mbh_{j,\text{sample}} \sim \calN_{\C}(\mbmu_{k^\star_j}, \mbC_{k^\star_j})$, which conveys statistical information about the channel of \ac{MT}~$j$.
By applying the \ac{SWMMSE} algorithm, the \ac{BS} can compute the precoders by utilizing the generated samples, cf.~\cite{TuFeKoJoUt23}.

\section{Enabling Variable Bit Lengths}
\label{sec:varbit}

In this Section, we discuss how the \ac{GMM}-based feedback scheme can be enabled to utilize variable bit lengths by only requiring a single \ac{GMM} of a certain size, of which \acp{GMM} with a smaller number of components can be obtained.
We investigate two different mixture reduction techniques, where given a \ac{GMM} with $K=2^B$ components, another \ac{GMM} with fewer components $K_S=2^{B_S}$, with, $K_S<K$, can be obtained.

There exist many methods to obtain a \ac{GMM} with fewer components starting from a \ac{GMM} with more components.
In this work, we analyze two types of mixture reduction techniques.
In particular, we use one \emph{merging} strategy which successively combines pairs of components with high similarity~\cite{Ru07} and one \emph{pruning} strategy, which simply discards components with a low contribution to the overall mixture~\cite{CrWiPaSv11}.

In this work, we focus our analysis on the merging technique from ~\cite{Ru07}.
Thereby, one replaces two mixture components with the smallest dissimilarity denoted by $\{\pi_a,\mbmu_a, \mbC_a\}$ and $\{\pi_b,\mbmu_b, \mbC_b\}$, by their moment preserving merge $\{\pi_m,\mbmu_m, \mbC_m\}$ \cite{Ru07}:
\begin{align}
    \pi_m &= \pi_a + \pi_b \nonumber \\
    \mbmu_m &= \tfrac{1}{\pi_m} (\pi_a \mbmu_a + \pi_b \mbmu_b) \label{eq:momentpreservemerge} \\
    \mbC_m &=  ( \tfrac{\pi_a}{\pi_m} \mbC_a +  \tfrac{\pi_b}{\pi_m} \mbC_b + \tfrac{\pi_a \pi_b}{\pi_m^2}(\mbmu_a - \mbmu_b)(\mbmu_a-\mbmu_b)^\herm)  \nonumber
\end{align}
The dissimilarity between two components is measured by
\begin{multline} \label{eq:dissimimeasure}
    d_{a,b} = \tfrac{1}{2} [ \pi_m \log \det(\mbC_m) \\
    - \pi_a \log \det(\mbC_a) - \pi_b \log \det(\mbC_b) ]
\end{multline}
which is an upper bound on the Kullback-Leibler divergence between the \acp{GMM} before and after the merge, cf. \cite{Ru07}.
The merging procedure is successively repeated until a \ac{GMM} of the desired size is obtained, i.e., when the number of \ac{GMM} components is reduced to~$2^{B_S}$.
After a \ac{GMM} of the desired size is obtained, the corresponding codebook needs to be constructed at the \ac{BS} in the same way as described in Section \ref{sec:coebookconstruction}.
Algorithm \ref{alg:merging} summarizes the merging procedure and the accompanying codebook construction.

\begin{algorithm}[t]
\captionsetup{font=footnotesize}
\footnotesize
\caption{Variable Bit Lengths Enabled Through Merging.}
\label{alg:merging}
    \begin{algorithmic}[1]
    \REQUIRE \ac{GMM} with $K=2^B$ components. Desired number of components $K_S=2^{B_S}$ with $K_S<K$. 
    \STATE $i = K$
    \COMMENT{track number of \ac{GMM} components}
    \REPEAT
    \STATE Calc. moment preserving merge of all pairs of components using~\eqref{eq:momentpreservemerge}.
    \STATE Find pair with smallest dissimilarity $d_{a,b}$, via \eqref{eq:dissimimeasure}.
    \STATE Replace this pair by their moment preserving merge.
    \STATE $i \gets i-1$
    \UNTIL{$i = K_S$}
    \STATE Partition $\mathcal{H}$ into $K_S$ disjoint sets $\mc{V}_k$ with \( k = 1, \dots, K_S \) according to the \ac{GMM} responsibilities, cf. \eqref{eq:gmmcb_stage_1}.
    \STATE Compute codebook \( \mc{Q} = \{ \mbQ_k \}_{k=1}^{K_S} \) by solving \eqref{eq:gmmcb_stage_2} for each entry.
    \end{algorithmic}
\end{algorithm}

\begin{algorithm}[t]
    \captionsetup{font=footnotesize}
    \footnotesize
    \caption{Variable Bit Lengths Enabled Through Pruning.}
    \label{alg:pruning}
        \begin{algorithmic}[1]
        \REQUIRE \ac{GMM} with $K=2^B$ components. Desired number of components $K_S=2^{B_S}$ with $K_S<K$. 
        \STATE Remove the $K-K_S$ components with smallest mixing coefficients $\pi_k$.
        \STATE Re-normalize remaining mixing coefficients to one.
        \STATE Discard codebook entries correspondingly and obtain \( \mc{Q}=\{ \mbQ_k \}_{k=1}^{K_S} \).
        \end{algorithmic}
\end{algorithm}

The pruning strategy simply removes $K-K_S$ components with the smallest mixing coefficient $\pi_k$ (see \eqref{eq:gmm_of_h}), and the mixing coefficients of the remaining components are re-normalized to one \cite{CrWiPaSv11}.
A main advantage of the pruning strategy is its low computational complexity.
In particular, it only requires traversing a list of sorted mixing coefficients.
In this case, an updated codebook $\mathcal{Q} = \{ \mbQ_k \}_{k=1}^{K_S} $ can simply be obtained by discarding the respective codebook entries which correspond to the removed \ac{GMM} components.
Algorithm \ref{alg:pruning} summarizes the pruning procedure.

\section{Reducing the Offloading Overhead}
\label{sec:reducedoverhead}

\begin{table}[t]
\renewcommand{\arraystretch}{1.3}
\begin{center}
\begin{tabular}{|l|c|c|c|}
\hline
\textbf{Name} & \textbf{Covariance Parameters} & \textbf{Example}\\ \hline
Full & \( \frac{1}{2} K N(N+1) \) & \( 8.4 \cdot 10^6 \) \\ \hline
Kronecker & $ \frac{1}{2}\Krx \Nrx(\Nrx+1)$  $+ \frac{1}{2}\Ktx \Ntx(\Ntx+1)$ & \( 9.0 \cdot 10^3 \) \\ \hline
Toeplitz & $ 2\Krx \Nrx$  $+ 4\Ktx \Ntx$ & \( 2.1 \cdot 10^3 \) \\ \hline
Circulant & $ \Krx \Nrx$  $+ \Ktx \Ntx$ & \( 5.7 \cdot 10^2 \) \\ \hline
\end{tabular}
\end{center}
\caption{Analysis of the number of covariance parameters of the (structured) \ac{GMM}.}
\label{tab:num_params}
\end{table}

To enable a \ac{MT} to compute feedback indices via \eqref{eq:ecsi_index}, the parameters of the \ac{GMM} need to be offloaded to the \ac{MT} upon entering the \ac{BS}' coverage area, cf. \cite{TuFeKoJoUt23}.
To reduce the offloading overhead model-based insights can be utilized to obtain structured covariances with fewer parameters.
For example, one can constrain the \ac{GMM} covariances to a Kronecker factorization of the form \( \covhk = \mbC_{\text{tx},k} \otimes \mbC_{\text{rx},k} \) which has no notable impact on the performance, cf. \cite{KoFeTuUt21J, TuFeKoJoUt23}.  
In this context, instead of fitting an unconstrained GMM with $N\times N$-dimensional covariances, a \ac{GMM} specific to the transmit side and another for the receive side is fitted. 
These transmit-side and receive-side \acp{GMM} have $N_{\text{tx}} \times N_{\text{tx}}$ and $N_{\text{rx}} \times N_{\text{rx}}$-dimensional covariances respectively, with $K_{\text{tx}}$ and $K_{\text{rx}}$ components. 
Subsequently, by computing all Kronecker products of the corresponding transmit-side and receive-side covariance matrices, a \( K = \Ktx\Krx \)-components \ac{GMM} with $N\times N$-dimensional covariances is obtained.

In this work, we investigate the incorporation of further structural features to the transmit- and receive-side \ac{GMM} covariance matrices imposed by the antenna structure at the transmitter or the receiver.
In case of a \ac{ULA}, it is common to assume a Toeplitz covariance matrix, which for a large number of antenna elements, is well approximated by a circulant matrix, cf., e.g.,~\cite{TuFeGrKoUt22}.
If a \ac{URA} is deployed, the structural assumptions result in block-Toeplitz matrices with Toeplitz blocks, or block-circulant matrices with circulant blocks, respectively \cite{TuFeGrKoUt22}.

In the assumed case of a \ac{URA} employed at the \ac{BS} with $\Ntxv$ vertical and $\Ntxh$ horizontal ($\Ntx=\Ntxv\Ntxh$) elements, the structured covariances can be expressed as $\mbC_{\text{tx},k} = \mbD^\herm \diag(\mbc_{\text{tx},k}) \mbD$, where on the one hand, when assuming a Toeplitz structure, $\mbD = \mbD_{\Ntxv} \otimes \mbD_{\Ntxh}$, where $\mbD_T$ contains the first $T$ columns of a $2T\times 2T$ \ac{DFT} matrix, and $\mbc_{\text{tx},k} \in \R_{+}^{4\Ntx}$ \cite{FeJoHuKoTuUt22, TuFeGrKoUt22}.
On the other hand, when assuming circular structure, we have $\mbD = \mbF_{\Ntxv} \otimes \mbF_{\Ntxh}$, where $\mbF_T$ is the $T\times T$ DFT-matrix, and $\mbc_{\text{tx},k} \in \R_{+}^{\Ntx}$.
At the \acp{MT}, we assume to have \acp{ULA}, thus $\mbD$ contains either the first $\Nrx$ columns of a $2\Nrx\times 2\Nrx$ \ac{DFT} matrix and correspondingly $\mbc_{\text{rx},k} \in \R_{+}^{2\Nrx}$  (Toeplitz), or $\mbD = \mbF_{\Nrx}$ and $\mbc_{\text{rx},k} \in \R_{+}^{\Nrx}$ (circulant).
With these structural constraints the \ac{GMM} covariances are fully determined by the vectors $\mbc_{\{\text{tx,rx}\},k}$.
Overall, these insights significantly reduce the offloading overhead, enable a reduced complexity in offline training, facilitate parallelization of the fitting process, and require fewer training samples since fewer parameters need to be learned, cf. \cite{FeJoHuKoTuUt22, TuFeGrKoUt22}.

Table \ref{tab:num_params} depicts the number of (structured) \ac{GMM} covariance parameters (accounting for symmetries).
We use exemplarily the simulation parameters of a configuration with $B=6$, $(\Ntx, \Nrx) = (32,16)$, and $(\Ktx, \Krx) = (16,4)$, which we consider in Section~\ref{sec:sim_results}.
We can observe that, structural constraints significantly reduce the offloading overhead.

\section{Discussion on the Enhanced Versatility}
\label{sec:discussion_universality}

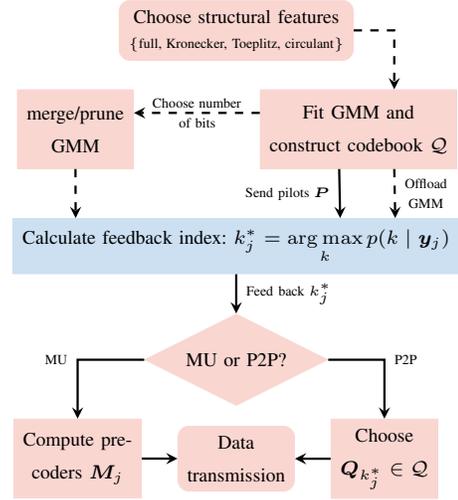
\begin{figure}[t]
    \centering
    \input{flowchart_feedback_scheme_structmerged}
    \caption{Flowchart of the improved versatile feedback scheme. Red (blue) colored nodes are processed at the \ac{BS} (\acp{MT}) and solid (dashed) arrows indicate online (offline) processing.}
    \label{fig:flowchart}
\end{figure}

The low-complexity \ac{GMM}-based feedback scheme from~\cite{TuFeKoJoUt23} exhibits great flexibility with respect to the transmission mode, i.e., either point-to-point \ac{MIMO} or \ac{MU-MIMO}, the supported range of \ac{SNR} values, the pilot configuration, and the selection of the precoding algorithm, but utilizes a \ac{GMM} with a predefined number of components $K=2^B$.
In principle, the \ac{BS} could offload many \acp{GMM} with different sizes, which would allow for different numbers of feedback bits.
However, the corresponding signaling overhead might be unaffordable in practice.
Therefore, in Section \ref{sec:varbit} we discussed different mixture reduction techniques and their corresponding codebook update procedures.
Particularly, given a \ac{GMM} with $K=2^B$ components, \acp{GMM} with fewer components $K_S=2^{B_S}$, i.e, $K_S<K$, can be obtained.
Accordingly, the scheme is enabled to support variable bit lengths by successively decreasing the number of components and thereby the number of feedback bits starting from $B$ bits ($B-1,B-2,\cdots$). 
Due to the simplicity of the pruning strategy, it can be straightforwardly implemented at the \acp{MT}. 
The merging strategy can be either conducted at the \acp{MT}, or to reduce their computational burden, the \ac{BS} could offload additional lists with the respective merging updates.
Moreover, the \ac{MT} only requires the \ac{GMM} and does not need to be aware of the codebook in order to compute the feedback via \eqref{eq:ecsi_index}.
Altogether, this allows for variable feedback bit lengths and eliminates the necessity of offloading a particularly trained \ac{GMM} for different feedback bit lengths and thereby improves the versatility of the feedback scheme.
Additionally, in Section~\ref{sec:reducedoverhead} we discussed how model-based insights can help to effectively reduce the offloading overhead.
\Cref{fig:flowchart} provides a flowchart of the enhanced \ac{GMM}-based feedback scheme.

\section{Simulation Results} \label{sec:sim_results}

We generate datasets for the \ac{UL} and \ac{DL} domain of the scenario outlined in Section \ref{sec:data_generation}: $\mathcal{H}^{\text{UL}}$ and $ \mathcal{H}^{\text{DL}} $.
The \ac{GMM} is fitted centrally at the \ac{BS} 
using the training set $\mathcal{H}=\mathcal{H}^{\text{UL}}$ which consists of $L = 20 \cdot 10^3$ samples, cf., \cite{TuFeKoJoUt23, TuKoBaXuUt21, fesl2021centralized, TuKoRiFeBaXuUt21, utschick2021}.
The following transmit strategies are always evaluated in the \ac{DL} domain using \( \mathcal{H}^{\text{DL}} \), comprised of $10^4$ channels.
The data samples are normalized to satisfy \( \expec[\|\mbh\|^2] = N = \Ntx \Nrx \). 
Additionally, we fix $\rho=1$, enabling the definition of the \ac{SNR} as \( \frac{1}{\sigma_n^2} \) for all \acp{MT}, specifically when $\sigma^2_j = \sigma^2_n, \forall j \in \mathcal{J}$.

In the single-user case, we will compare to codebook based approaches utilizing Lloyd's clustering algorithm, cf. \cite{LaYoCh04, TuFeKoJoUt23}.
In case of \ac{MU-MIMO} systems, we will either use directional information extracted from the Lloyd codebooks or we will use the random codebook approach, cf. \cite{TuFeKoJoUt23, RaJi08}.
With these approaches, prior to codebook entry selection, the channel needs to be estimated.
To this end, we consider three different channel estimators, which are briefly explained in the following.
The recently proposed \ac{GMM}-based channel estimator \( \hhat_{\text{GMM}} \), from \cite{KoFeTuUt21J} is one of them and utilizes the same \ac{GMM} as found in Section~\ref{sec:proposedscheme} and calculates a convex combination of per-component \ac{LMMSE} estimates, cf. \cite{TuFeKoJoUt23, KoFeTuUt21J}.
This estimator is proven to asymptotically converge to the optimal conditional mean estimator as the number of components $K$ is increased, cf. \cite{KoFeTuUt21J}.
Another baseline is the \ac{LMMSE} estimator $\hhat_{\text{LMMSE}}$, where the sample covariance matrix is constructed given the set $\mathcal{H}$, cf. \cite{TuFeKoJoUt23, fesl2021centralized, KoFeTuUt21J}.
Lastly, we consider a compressive sensing estimation approach $\hhat_{\text{OMP}}$ utilizing the \ac{OMP}, cf.~\cite{TuFeKoJoUt23, AlLeHe15}.

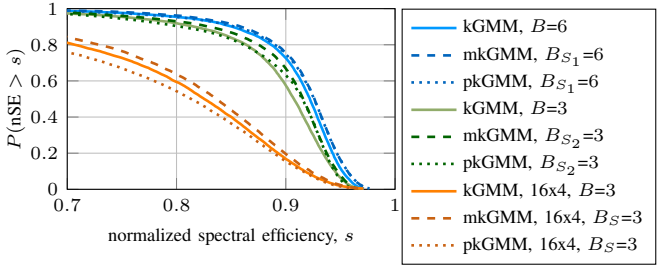
\begin{figure}[tb]
    \centering
        \input{fig_cb_merged_calc_pruned_16x4_32x16}
    \caption{Empirical \acp{cCDF} of the normalized spectral efficiencies for two setups using either mixture reduction techniques or the direct fitting approach \textbf{assuming perfect \ac{CSI}}. Setup~A: $\Ntx=32$ ($\Ntxh=8, \Ntxv=4$), $\Nrx=16$, and $\text{SNR}=\SI{10}{dB}$. Setup~B: $\Ntx=16$ ($\Ntxh=4, \Ntxv=4$), $\Nrx=4$, and $\text{SNR}=\SI{0}{dB}$.}
    \vspace{-0.25cm}
    \label{fig:mergeprune_twosettings}
\end{figure}

\subsection{Point-to-point MIMO}

As performance measure, we utilize the \ac{nSE}, where the spectral efficiency achieved with a certain transmit strategy is normalized by the optimal water-filling solution, cf. \cite{Love}.
The empirical \ac{cCDF} $P(\text{nSE}>s)$ for the normalized spectral efficiency represents the empirical probability that the \ac{nSE} surpasses a given value $s$.

\begin{figure}[tb]
    \centering
        \input{fig_cb_toep_circ_16x4_32x16}
    \caption{Empirical \acp{cCDF} of the normalized spectral efficiencies for two setups using differently structured \acp{GMM} \textbf{assuming perfect \ac{CSI}}. Setup~A: $\Ntx=32$ ($\Ntxh=8, \Ntxv=4$), $\Nrx=16$, and $\text{SNR}=\SI{10}{dB}$, with $B=6$. Setup~B: $\Ntx=16$ ($\Ntxh=4, \Ntxv=4$), $\Nrx=4$, and $\text{SNR}=\SI{0}{dB}$, with $B=3$.}
    \label{fig:circtoep_twosettings}
    \vspace{-0.25cm}
\end{figure}
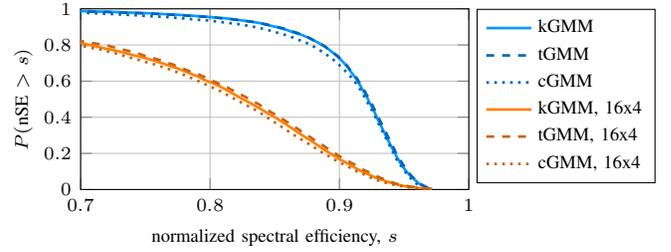

In \Cref{fig:mergeprune_twosettings}, we denote by ``Setup A'' a system with $\Ntx=32$ $(\Ntxh=8, \Ntxv=4)$, $\Nrx=16$, and $\text{SNR}=\SI{10}{dB}$, and with ``Setup B'' a system with $\Ntx=16$ $(\Ntxh=4, \Ntxv=4)$, $\Nrx=4$, and $\text{SNR}=\SI{0}{dB}$.
In the following, we consider the Kronecker \ac{GMM} (``kGMM'') with full transmit- and receive-side \acp{GMM} and assume perfect \ac{CSI}, i.e., the feedback information is determined via \eqref{eq:pcsi_index}.
For ``Setup~A'' we fit a \ac{GMM} with $K=256$ $(B=8$, $\Ktx=32$, and $\Krx=8)$ components and first reduce to a \ac{GMM} with $K_{S_1}=64$ $(B_{S_1}=6)$ components and continue the reduction to another one with $K_{S_2}=8$ $(B_{S_2}=3)$ components, by either applying the merging strategy (``mkGMM'') or the pruning strategy (``pkGMM'').
Moreover, with ``kGMM'' we denote the case of directly fitting a \ac{GMM} with $64$ $(B=6$, $\Ktx=16$, and $\Krx=4)$, or with $8$ $(B=3$, $\Ktx=4$, and $\Krx=2)$ components.
We can observe, that both mixture reduction approaches not only enable the feedback scheme with variable bit lengths, but at the same time provide at least a similar performance as the respective direct fitting approaches.
In case of ``Setup B'', we fit a \ac{GMM} with $K=256$ components and reduce to one with $K_S=8$ $(B_S=3)$ components and compare to a \ac{GMM} directly fitted with $8$ $(\Ktx=4$, and $\Krx=2)$ components.
In this setup, the merging strategy performs best, whereas the low-complexity pruning strategy yields a slightly worse performance as compared to the direct fitting approach.
Altogether, the applied mixture reduction techniques enable variable bit lengths and in some cases even improve the performance as a consequence of the enhanced clustering ability, cf. \cite{He10}.

In \Cref{fig:circtoep_twosettings}, consider the same two setups and denote  by ``tGMM'' or ``cGMM'' Kronecker \acp{GMM} constructed by (block) Toeplitz or (block) circulant transmit- and receive-side \acp{GMM}, respectively, with $64$ components $(B=6$, $\Ktx=16$, and $\Krx=4)$ in case of ``Setup A'', and $8$ components $(B=3$, $\Ktx=4$, and $\Krx=2)$ in case of ``Setup B''.
We can observe, that although the Toeplitz approximation (``tGMM'') drastically reduces the offloading overhead, it achieves the same performance as the Kronecker \ac{GMM} with full transmit- and receive-side \acp{GMM} (``kGMM'').
The circulant approximation (``cGMM'') with even less offloading overhead, slightly degrades the performance.

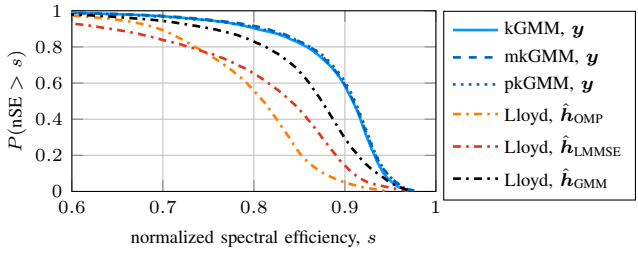
\begin{figure}[tb]
    \centering
        \input{fig_p2p_eccdf_merged_pruned}
    \caption{Empirical \acp{cCDF} of the normalized spectral efficiencies using either mixture reduction techniques $(B=8$ to $B_S=6)$ or the direct fitting approach $(B=6)$ for a system with $\Ntx=32$, $\Nrx=16$, $\text{SNR}=\SI{15}{dB}$, and $n_p=4$ pilots.}
    \label{fig:mergeprune_p2p_imperfect}
    \vspace{-0.25cm}
\end{figure}

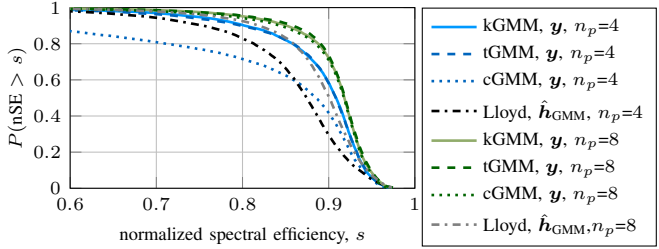
\begin{figure}[tb]
    \centering
        \input{fig_p2p_eccdf_structured}
    \caption{Empirical \acp{cCDF} of the normalized spectral efficiencies using differently structured \acp{GMM} for a system with $\Ntx=32$, $\Nrx=16$, $\text{SNR}=\SI{15}{dB}$, different number of pilots $n_p$, and $B=6$ bits.}
    \label{fig:circtoep_p2p_imperfect}
    \vspace{-0.5cm}
\end{figure}

Nevertheless, assuming perfect \ac{CSI} at the \ac{MT} during the online phase is impractical.
In the subsequent discussions, we consider imperfect \ac{CSI}, i.e., systems characterized by low pilot overhead $(n_p \leq \Ntx)$.
With ``Lloyd, \{$\hhat_{\text{GMM}}$, $\hhat_{\text{OMP}}, \hhat_{\text{LMMSE}}$\}'' we depict the conventional approaches which first estimate the channel and then determine the feedback information, cf. \cite{TuFeKoJoUt23}.
In the remainder, the conventional approaches always utilize the same number of feedback bits of interest, i.e., $B_S$ bits if merging or pruning is considered, or $B$ bits if not.

In \Cref{fig:mergeprune_p2p_imperfect}, we simulate a setup with $\Ntx=32$ $(\Ntxh=8, \Ntxv=4)$, $\Nrx=16$, $\text{SNR}=\SI{15}{dB}$, and $n_p=4$ pilots.
We again consider a \ac{GMM} with $K=256$ $(B=8$, $\Ktx=32$, and $\Krx=8)$ components and reduce to a \ac{GMM} with $K_S=64$ $(B_S=6)$ components, by applying the merging strategy (``mkGMM'') or the pruning strategy (``pkGMM'').
We can observe, that the \ac{GMM}-based feedback approach (``kGMM, $\mby$''), which circumvents explicit channel estimation, is superior as compared to the baselines and provides a higher robustness against \ac{CSI} imperfections.
The proposed merging and pruning strategies exhibit a similar robustness.

In \Cref{fig:circtoep_p2p_imperfect}, we have $\Ntx=32$ $(\Ntxh=8, \Ntxv=4)$, $\Nrx=16$, $\text{SNR}=\SI{15}{dB}$, and $n_p \in \{4, 8\}$ pilots and depict differently structured \acp{GMM} with $K=64$ $(\Ktx=16$, and $\Krx=4)$ components.
The Toeplitz approximation (``tGMM'') with a reduced number of parameters achieves the same performance as compared to the Kronecker \ac{GMM} with full transmit- and receive-side \acp{GMM} (``kGMM''), irrespective of the number of pilots.
With $n_p=4$, the proposed approach ``tGMM'' attains the same performance as the conventional Lloyd approach with twice as much pilots ($n_p=8$).
This shows, the great potential of the enhanced \ac{GMM}-based feedback scheme in systems with reduced pilot overhead.
Interestingly, in case of the circulant approximation (``cGMM'') a performance degradation can be observed if $n_p=4$.
It seems that the expressivity of the circulant \ac{GMM} with a very low number of parameters is too restrictive to be applied for systems with a few pilots only.
Since the Toeplitz approach provides a higher robustness, we will restrict our further analysis to it in the remainder.

\subsection{Multi-user MIMO}

For the multi-user scenario, we utilize the sum-rate as performance metric, cf. \cite{TuFeKoJoUt23}.
The presented results are based on $2{,}500$ constellations, where for each constellation, we randomly select $J$ \acp{MT} from our evaluation set \( \mathcal{H}^{\text{DL}} \).
We employ the empirical \ac{cCDF} $P(\text{SR}>s)$ of the sum-rate to illustrate the empirical probability of the \ac{SR} exceeding a given value $s$.
Although the \ac{GMM}-based feedback scheme can be used in combination with non-iterative precoding algorithms, in the following we restrict our analysis to the iterative \ac{WMMSE}  and to the \ac{SWMMSE} in order to design the precoders, since the performance with these methods is generally better, cf. \cite{TuFeKoJoUt23}.

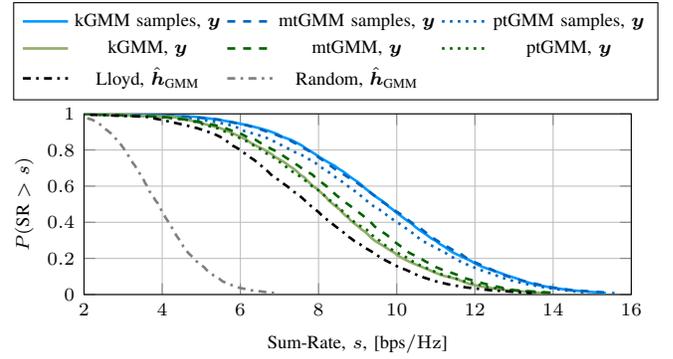
\begin{figure}[tb]
    \centering
    \input{fig_mumimo_eccdf_iter}
    \caption{Empirical \acp{cCDF} of the sum-rate 
    using either mixture reduction techniques $(B=8$ to $B_S=6)$ combined with Toeplitz structured \acp{GMM} or the direct fitting approach $(B=8)$ when the iterative \ac{WMMSE} or the \ac{SWMMSE} are utilized, for a system with $\Ntx=16$, $\Nrx=4$, $J=4$ \acp{MT}, $\text{SNR}=\SI{5}{dB}$, and $n_p=8$ pilots.}
    \label{fig:mumimonp8eccdf}
    \vspace{-0.5cm}
\end{figure}

In the subsequent discussion, we refer with ``GMM, $\mby$'' to the case where the observations $\mby_j$ are used at each \ac{MT} $j$ to determine a feedback index utilizing the \ac{GMM} feedback encoding approach, cf. \eqref{eq:ecsi_index_j}.
The channel of each \ac{MT} is then represented by the subspace information extracted from the \ac{GMM} codebook, cf. Section \ref{sec:mumimo}.
For the sake of clarity in the legend, we omit the index $j$.
With ``\{Lloyd, Random\}, \{$\hhat_{\text{GMM}}$, $\hhat_{\text{OMP}}, \hhat_{\text{LMMSE}}$\}'', we denote the cases where the channel is estimated at each \ac{MT} and subsequently the feedback information is determined using either the directional Lloyd codebook or the random codebook approach, cf. \cite{TuFeKoJoUt23}.
These methods use the iterative \ac{WMMSE}, cf. \cite[Algorithm~1]{HuCaYuQiXuYuDi21}.
Furthermore, with ``GMM samples, $\mby$'' we refer to the case, where we generate samples 
representing the distribution of each \ac{MT} using the \ac{GMM} and feed these samples to the SWMMSE algorithm, cf. Section~\ref{sec:mumimo}.
The maximum number of iterations is $I_{\max}=300$ for both approaches.

In the remainder, we combine both, the Toeplitz approximation and the mixture reduction techniques in a setup with $\Ntx=16$ ($\Ntxh=4, \Ntxv=4$) and $\Nrx=4$, and $J=4$ users.
In particular, we fit a Kronecker \ac{GMM} constructed by (block) Toeplitz transmit- and receive-side \acp{GMM}, with $K=256$ components and reduce to a \ac{GMM} with $K_S=64$ ($B_S=6$) components denoted by ``mtGMM''and ``ptGMM'', and compare to a \ac{GMM} directly fitted with $64$ ($\Ktx=16$, and $\Krx=4$) components with full transmit- and receive-side \acp{GMM} (``kGMM'').
In \Cref{fig:mumimonp8eccdf}, with $\text{SNR}=\SI{5}{dB}$ and $n_p=8$ pilots, we can see that the pruning and merging strategy in combination with the Toeplitz structured covariances in case of the directional (subspace) approach, i.e., ``ptGMM, $\mby$'' and ``mtGMM, $\mby$'', perform similarly or even better than the ``kGMM, $\mby$'' and outperform the conventional methods ``\{Lloyd, Random\}, $\hhat_{\text{GMM}}$'', by far.
With the generative modeling approach, the merging method ``mtGMM samples, $\mby$'' performs equally well as ``kGMM samples, $\mby$'', whereas the pruning method ``ptGMM samples, $\mby$'' performs slightly worse.

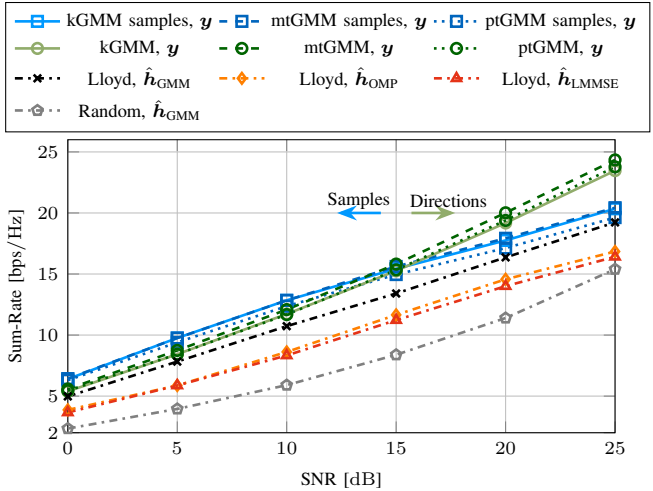
\begin{figure}[tb]
    \centering
    \input{fig_mumimo_oversnr}
    \caption{The average sum-rate over the \ac{SNR} using either mixture reduction techniques $(B=8$ to $B_S=6)$ combined with Toeplitz structured \acp{GMM} or the direct fitting approach $(B=6)$ when the iterative \ac{WMMSE} or the \ac{SWMMSE} are employed, for a system with $\Ntx=16$, $\Nrx=4$, $J=4$ \acp{MT}, and $n_p=8$ pilots.}
    \label{fig:mumimonp8sroversnr}
    \vspace{-0.4cm}
\end{figure}

A similar behaviour is present in \Cref{fig:mumimonp8sroversnr}, where we still consider the same setting with $n_p=8$ pilots, but vary the \ac{SNR} and depict the sum-rate averaged over all constellations.
Interestingly, with an increasing \ac{SNR}, the gap between ``mtGMM, $\mby$'' and ``kGMM, $\mby$'' increases.
Similarly, the gap between ``ptGMM samples, $\mby$'' and ``kGMM samples, $\mby$'' increases with larger \ac{SNR} values.
As reported in \cite{TuFeKoJoUt23}, adopting the generative modeling approach for joint precoder design is advantageous in scenarios characterized by low to moderate \ac{SNR} levels, and for larger \ac{SNR} values the directional approach is beneficial.
This property can be similarly observed if the merging or pruning procedures are applied, and is illustrated by the arrows in \Cref{fig:mumimonp8sroversnr}. 
Altogether, the enhanced \ac{GMM}-based feedback scheme exhibits superior performance compared to the baselines in a multi-user system with reduced pilot overhead.

\section{Conclusion}

In this work, we enhanced the versatile \ac{GMM}-based feedback from \cite{TuFeKoJoUt23} scheme by enabling variable bit lengths and incorporating further model-based structural constraints to reduce the offloading overhead.
To this end, we analyzed two mixture reduction techniques, i.e., one pruning and one merging strategy and found out, that both approaches enable the scheme with variable bit lengths and allow for a trade-off between complexity and performance.
Moreover, numerical results showed the remarkable performance of the combination of mixture reduction techniques with structured \acp{GMM}.

\bibliographystyle{IEEEtran}
\bibliography{IEEEabrv,biblio}
\end{document}

%% file: miko.tex
\usepackage{mathtools}	
\usepackage{amssymb}
\usepackage{amsthm}

\usepackage[utf8]{inputenc}

\usepackage{ifthen}

\usepackage{microtype}

\usepackage{caption}

\usepackage{bm}

\usepackage[draft,nomargin,marginclue,inline]{fixme}
\makeatletter
\renewcommand*\FXLayoutMarginClue[3]{%
  \marginpar[%
  \raggedleft\@fxuseface{margin}\textcolor{red}{\ignorespaces $ \Rightarrow $}]{%
    \raggedright\@fxuseface{margin}\textcolor{red}{\ignorespaces $ \Leftarrow $}}}
\makeatother	

\usepackage{tumcolor}

\usepackage{pgfplotstable}	

\pgfplotsset{
	discard if/.style 2 args={
        x filter/.append code={
            \edef\tempa{\thisrow{#1}}
            \edef\tempb{#2}
            \ifx\tempa\tempb
                
            \fi
        }
    },
    discard if not/.style 2 args={
        x filter/.append code={
            \edef\tempa{\thisrow{#1}}
            \edef\tempb{#2}
            \ifx\tempa\tempb
            \else
                
            \fi
        }
    }
}

\usepackage{cite}

\usepackage[acronym,shortcuts]{glossaries}
\newacronym{cnn}{CNN}{convolutional neural network}
\newacronym{ula}{ULA}{uniform linear array}

\usepackage{cleveref}

\usepackage{enumitem}

\tikzset{algorithm1/.style={mark options={solid},color=TUMBeamerBlue, line width=\lineWidth, mark=square, dashed}}

\pgfplotscreateplotcyclelist{miko}{%
{color=TUMBeamerRed, dash pattern=on 5pt off 1pt on 1pt off 1pt, line width=1.5pt},%
{color=TUMBeamerOrange, dash pattern=on 5pt off 2pt, line width=1.5pt,
mark=o},%
{color=black, dash pattern=on 5pt off 2pt on 3pt off 2pt, line width=1.5pt},%
{color=TUMDarkerBlue, line width=1.5pt},%
{color=black, dotted, line width=1.5pt},%
{color=TUMBeamerGreen, dotted, line width=1.5pt},%
{color=TUMBeamerYellow, dotted, line width=1.5pt},%
{color=TUMBeamerDarkRed, dotted, line width=1.5pt},%
{color=TUMBeamerLightGreen, dotted, line width=1.5pt},%
{color=TUMIvory, dotted, line width=1.5pt},%
{color=TUMLightBlue, dotted, line width=1.5pt},%
}

\DeclareMathOperator*{\argmax}{arg\,max}

\DeclareMathOperator{\diag}{diag}

\DeclareMathOperator{\expec}{E}

\DeclareMathOperator{\vect}{vec}

\newcommand*{\mc}[1]{\mathcal{#1}}	

\newcommand{\calN}{\mathcal{N}}
\newcommand{\calO}{\mathcal{O}}

\newcommand*{\C}{\mathbb{C}}

\newcommand*{\R}{\mathbb{R}}

\newcommand{\herm}{{\operatorname{H}}}
\newcommand{\tp}{{\operatorname{T}}}

\definecolor{myblue}{RGB}{30, 100, 200}

\newlength{\leftstackrelawd}
\newlength{\leftstackrelbwd}
\def\leftstackrel#1#2{\settowidth{\leftstackrelawd}%
	{${{}^{#1}}$}\settowidth{\leftstackrelbwd}{$#2$}%
	\addtolength{\leftstackrelawd}{-\leftstackrelbwd}%
	\leavevmode\ifthenelse{\lengthtest{\leftstackrelawd>0pt}}%
	{\kern-.5\leftstackrelawd}{}\mathrel{\mathop{#2}\limits^{#1}}}

\newcommand{\mbA}{\bm{A}}

\newcommand{\mbC}{\bm{C}}
\newcommand{\mbD}{\bm{D}}

\newcommand{\mbF}{\bm{F}}

\newcommand{\mbH}{\bm{H}}

\newcommand{\mbM}{\bm{M}}
\newcommand{\mbN}{\bm{N}}

\newcommand{\mbP}{\bm{P}}
\newcommand{\mbQ}{\bm{Q}}

\newcommand{\mbT}{\bm{T}}

\newcommand{\mbX}{\bm{X}}
\newcommand{\mbY}{\bm{Y}}

\newcommand{\mbc}{\bm{c}}

\newcommand{\mbh}{\bm{h}}

\newcommand{\mbn}{\bm{n}}

\newcommand{\mbp}{\bm{p}}

\newcommand{\mbs}{\bm{s}}

\newcommand{\mbx}{\bm{x}}
\newcommand{\mby}{\bm{y}}

\newcommand{\mbSigma}{{\bm{\Sigma}}}

\newcommand{\mbmu}{{\bm{\mu}}}

\newcommand{\mbzero}{{\bm{0}}}

\newcommand{\hhat}{\hat{\mbh}}

\newcommand{\covhi}{\mbC_i}
\newcommand{\covhk}{\mbC_k}
\newcommand{\meanhi}{\mbmu_i}
\newcommand{\meanhk}{\mbmu_k}

%% file: flowchart_feedback_scheme_structmerged.tex
    \resizebox{0.7\columnwidth}{!}{
\begin{tikzpicture}[node distance=1.3cm]
	\tikzstyle{startstop} = [rectangle, rounded corners, minimum width=0.7cm, minimum height=0.7cm,text centered, draw=TUMBeamerRed!20, fill=TUMBeamerRed!20]
	\tikzstyle{io} = [trapezium, trapezium left angle=70, trapezium right angle=110, minimum width=1cm, minimum height=0.7cm, text centered, draw=white, fill=blue!30]
	\tikzstyle{process} = [rectangle, minimum width=0.7cm, minimum height=0.7cm, text centered, draw=TUMBlue!20, fill=TUMBlue!20]
	\tikzstyle{process_bs} = [rectangle, minimum width=1cm, minimum height=1cm, text centered, draw=TUMBeamerRed!20, fill=TUMBeamerRed!20]
	\tikzstyle{decision} = [diamond, minimum width=0cm, minimum height=0cm, text centered, draw=TUMBeamerRed!20, fill=TUMBeamerRed!20, aspect=2]
	\tikzstyle{arrow} = [thick,->,>=stealth]
	
	\node[align=center] (struct) [startstop] {\scriptsize Choose structural features\\[-3pt] \tiny \{full, Kronecker, Toeplitz, circulant\}};
	\node[align=center] (prune) [process_bs, below left=0.75cm and 1.3cm of struct.center] {\scriptsize merge/prune \\ \scriptsize GMM};
	\node[align=center] (start) [process_bs, below right=0.75cm and 0.3cm of struct.center] {\scriptsize Fit GMM and\\ \scriptsize construct codebook $\mathcal{Q}$};
	\draw [arrow,dashed] (struct) -| node[anchor=north east] {} ([shift=({-0.8cm,0})]start.north east);
	\draw [arrow,dashed] ([shift=({0,-0.3cm})]start.north west) -- node[align=center,anchor=center] {\tiny Choose number \\[-5pt] \tiny of bits} ([shift=({0,-0.3cm})]prune.north east);

	\node (feedback) [process, below=2cm of struct] {\scriptsize Calculate feedback index: $k^*_j = \argmax\limits_{k} p(k\mid \bm{y}_j)$};
	\draw [arrow,dashed] (prune.south) -- node[anchor=north east] {} ([shift=({0.8cm,0})]feedback.north west);
	
	
	\draw [arrow,align=center, dashed] ([shift=({-0.8cm,0})]start.south east) -- node[anchor=west] {\tiny Offload \\[-5pt] \tiny GMM} ([shift=({-0.82cm,0})]feedback.north east);
	\draw [arrow,align=center] ([shift=({-1.5cm,0})]start.south east) -- node[anchor=east] {\tiny Send pilots $\mbP$} ([shift=({-1.52cm,0})]feedback.north east);

	\node (dec1) [decision, below=0.5cm of feedback] {\scriptsize MU or P2P?};
	\draw [arrow] ([shift=({0,0})]feedback) -- node[anchor=west] {\tiny Feed back $k_j^*$} ([shift=({0,0})]dec1);
	\node (P2P) [process_bs, below right=0.75cm and 1.2cm of dec1.center,align=center] {\scriptsize Choose \\ \scriptsize $\bm{Q}_{k_j^*} \in \mathcal{Q}$};
	\draw [arrow] (dec1) -| node[anchor=west] {\tiny P2P} (P2P);
	\node (MU) [process_bs, below left=0.75cm and 1.2cm of dec1.center,align=center] {\scriptsize Compute pre-\\ \scriptsize coders $\bm{M}_j$};
	\draw [arrow] (dec1) -| node[anchor=east] {\tiny MU} (MU);
	\node (final) [startstop,  below=0.84cm of dec1.center,align=center] {\scriptsize Data \\\scriptsize transmission};
	\draw [arrow] (MU) -- node[anchor=east] {} (final);
	\draw [arrow] (P2P) -- node[anchor=east] {} (final);
\end{tikzpicture}
}

%% file: fig_cb_merged_calc_pruned_16x4_32x16.tex
    \begin{tikzpicture}
        \begin{axis}[
            height=\smallplotheight,
            width=0.925*\smallplotwidth,
            legend pos= north east,
            legend style={font=\scriptsize, at={(1.02, 1.0)}, anchor=north west, legend columns=1, cells={anchor=west}},
            label style={font=\scriptsize},
            y label style={at={(-0.1,0.5)}},
            tick label style={font=\scriptsize},
            title style={align=left, font=\scriptsize},
            xmin=0.7,
            xmax=1,
            ymin=0,
            ymax=1,
            xlabel={normalized spectral efficiency, $s$},
            ylabel={$P(\text{nSE}>s)$},
            grid=both,
        ]
            \addplot[gmmmrgmrg,TUMBeamerBlue]
                table[x=GMM_PGD_UL,y=oneminy_label,col sep=comma] {gmm_feedback_variableB/B6B6/all_se_rel_opt_cdf_gap40_nc64_SNR10dB_ntrain20000.csv};
                    \addlegendentry{kGMM, $B$=$6$};

            \addplot[gmmmrgmrg,TUMBlue, dashed]
                table[x=GMM_merged_cb_calc,y=oneminy_label,col sep=comma] {gmm_feedback_variableB/B8B6/all_se_rel_opt_merged_cdf_gap40_nc64_SNR10dB_ntrain20000.csv};
                    \addlegendentry{mkGMM, $B_{S_1}$=$6$};

            \addplot[gmmmrgmrg,TUMBlue, dotted]
                table[x=GMM_pruned_cb_pruned,y=oneminy_label,col sep=comma] {gmm_feedback_variableB/B8B6/all_se_rel_opt_pruned_cdf_gap40_nc64_SNR10dB_ntrain20000.csv};
                    \addlegendentry{pkGMM, $B_{S_1}$=$6$};

            \addplot[gmmmrgmrg,TUMBeamerGreen]
                table[x=GMM_PGD_UL,y=oneminy_label,col sep=comma] {gmm_feedback_variableB/B3B3/all_se_rel_opt_cdf_gap40_nc8_SNR10dB_ntrain20000.csv};
                    \addlegendentry{kGMM, $B$=$3$};

            \addplot[gmmmrgmrg,ourdarkgreen, dashed]
                table[x=GMM_merged_cb_calc,y=oneminy_label,col sep=comma] {gmm_feedback_variableB/B8B3/all_se_rel_opt_merged_cdf_gap40_nc8_SNR10dB_ntrain20000.csv};
                    \addlegendentry{mkGMM, $B_{S_2}$=$3$};

            \addplot[gmmmrgmrg,ourdarkgreen, dotted]
                table[x=GMM_pruned_cb_pruned,y=oneminy_label,col sep=comma] {gmm_feedback_variableB/B8B3/all_se_rel_opt_pruned_cdf_gap40_nc8_SNR10dB_ntrain20000.csv};
                    \addlegendentry{pkGMM, $B_{S_2}$=$3$};                    
                    


            \addplot[gmmmrgmrg,TUMBeamerOrange]
                table[x=GMM_PGD_UL,y=oneminy_label,col sep=comma] {gmm_feedback_variableB/B3B3_16x4/all_se_rel_opt_cdf_gap40_nc8_SNR0dB_ntrain20000.csv};
                    \addlegendentry{kGMM, 16x4, $B$=$3$};
            \addplot[gmmmrgmrg,ourdarkorange, dashed]
                table[x=GMM_merged_cb_calc,y=oneminy_label,col sep=comma] {gmm_feedback_variableB/B8B3_16x4/all_se_rel_opt_merged_cdf_gap40_nc8_SNR0dB_ntrain20000.csv};
                    \addlegendentry{mkGMM, 16x4, $B_{S}$=$3$};
            \addplot[gmmmrgmrg,ourdarkorange,dotted]
                table[x=GMM_pruned_cb_pruned,y=oneminy_label,col sep=comma] {gmm_feedback_variableB/B8B3_16x4/all_se_rel_opt_pruned_cdf_gap40_nc8_SNR0dB_ntrain20000.csv};
                    \addlegendentry{pkGMM, 16x4, $B_{S}$=$3$};

            

        \end{axis}
    \end{tikzpicture}

%% file: fig_cb_toep_circ_16x4_32x16.tex
    \begin{tikzpicture}
        \begin{axis}[
            height=\smallplotheight,
            width=1.05*\smallplotwidth,
            legend pos= north east,
            legend style={font=\scriptsize, at={(1.02, 1.0)}, anchor=north west, legend columns=1, cells={anchor=west}},
            label style={font=\scriptsize},
            y label style={at={(-0.1,0.5)}},
            tick label style={font=\scriptsize},
            title style={align=left, font=\scriptsize},
            xmin=0.7,
            xmax=1,
            ymin=0,
            ymax=1,
            xlabel={normalized spectral efficiency, $s$},
            ylabel={$P(\text{nSE}>s)$},
            grid=both,
        ]

            \addplot[gmmmrgmrg,TUMBeamerBlue]
                table[x=GMM_PGD_UL,y=oneminy_label,col sep=comma] {gmm_feedback_variableB/B6B6/all_se_rel_opt_cdf_gap40_nc64_SNR10dB_ntrain20000.csv};
                    \addlegendentry{kGMM};

            \addplot[gmmmrgmrg,TUMBlue, dashed]
                table[x=GMM_PGD_UL,y=oneminy_label,col sep=comma] {gmm_feedback_variableB/B6toep/all_se_rel_opt_cdf_gap40_nc64_SNR10dB_ntrain20000.csv};
                    \addlegendentry{tGMM};

            \addplot[gmmmrgmrg,TUMBlue, dotted]
                table[x=GMM_PGD_UL,y=oneminy_label,col sep=comma] {gmm_feedback_variableB/B6circ/all_se_rel_opt_cdf_gap40_nc64_SNR10dB_ntrain20000.csv};
                    \addlegendentry{cGMM};




            \addplot[gmmmrgmrg,TUMBeamerOrange]
                table[x=GMM_PGD_UL,y=oneminy_label,col sep=comma] {gmm_feedback_variableB/B3B3_16x4/all_se_rel_opt_cdf_gap40_nc8_SNR0dB_ntrain20000.csv};
                    \addlegendentry{kGMM, 16x4};

            \addplot[gmmmrgmrg,ourdarkorange, dashed]
                table[x=GMM_PGD_UL,y=oneminy_label,col sep=comma] {gmm_feedback_variableB/B3toep_16x4/all_se_rel_opt_cdf_gap40_nc8_SNR0dB_ntrain20000.csv};
                    \addlegendentry{tGMM, 16x4};

            \addplot[gmmmrgmrg,ourdarkorange, dotted]
                table[x=GMM_PGD_UL,y=oneminy_label,col sep=comma] {gmm_feedback_variableB/B3circ_16x4/all_se_rel_opt_cdf_gap40_nc8_SNR0dB_ntrain20000.csv};
                    \addlegendentry{cGMM, 16x4};
        \end{axis}
    \end{tikzpicture}

%% file: fig_p2p_eccdf_merged_pruned.tex
    \begin{tikzpicture}
        \begin{axis}[
            height=\smallplotheight,
            width=\smallplotwidth,
            legend pos= north east,
            legend style={font=\scriptsize, at={(1.02, 1.0)}, anchor=north west, legend columns=1, cells={anchor=west}},
            label style={font=\scriptsize},
            tick label style={font=\scriptsize},
            y label style={at={(-0.1,0.5)}},
            xmin=0.6,
            xmax=1,
            ymin=0.0,
            ymax=1,
            xlabel={normalized spectral efficiency, $s$},
            ylabel={$P(\text{nSE}>s)$},
            grid=both,
            yshift=-3.5cm,
        ]
        
            \addplot[gmmpgdulfromy, TUMBeamerBlue]
                table[x=GMM_PGD_UL_from_y,y=oneminy_label,col sep=comma] {gmm_feedback/P4/UL_all_se_rel_opt_cdf_gap40_nc64_SNR15dB_ntrain20000.csv};
                \addlegendentry{kGMM, $\mby$};
            \addplot[gmmpgdulfromy, TUMBlue, dashed]
                table[x=GMM_PGD_UL_from_y,y=oneminy_label,col sep=comma] {gmm_feedback_variableB/P4_B8B6merged/UL_all_se_rel_opt_cdf_gap40_nc64_SNR15dB_ntrain20000.csv};
                \addlegendentry{mkGMM, $\mby$};

            \addplot[gmmpgdulfromy, TUMBlue, dotted]
                table[x=GMM_PGD_UL_from_y,y=oneminy_label,col sep=comma] {gmm_feedback_variableB/P4_B8B6pruned/UL_all_se_rel_opt_cdf_gap40_nc64_SNR15dB_ntrain20000.csv};
                \addlegendentry{pkGMM, $\mby$};


            \addplot[lloydpgdulomp,dash dot]
                table[x=Lloyd_PGD_UL_omp,y=oneminy_label,col sep=comma] {gmm_feedback/P4/UL_all_se_rel_opt_cdf_gap40_nc64_SNR15dB_ntrain20000.csv};
                \addlegendentry{\pltlloydpgdulomp};
            \addplot[lloydpgdulscov,dash dot]
                table[x=Lloyd_PGD_UL_scovUL,y=oneminy_label,col sep=comma] {gmm_feedback/P4/UL_all_se_rel_opt_cdf_gap40_nc64_SNR15dB_ntrain20000.csv};
                \addlegendentry{\pltlloydpgdulscov};
            \addplot[lloydpgdulgmmulall,dash dot]
                table[x=Lloyd_PGD_UL_gmmULfull,y=oneminy_label,col sep=comma] {gmm_feedback/P4/UL_all_se_rel_opt_cdf_gap40_nc64_SNR15dB_ntrain20000.csv};
                \addlegendentry{\pltlloydpgdulgmmulall};

        \end{axis}
    \end{tikzpicture}

%% file: fig_p2p_eccdf_structured.tex
    \begin{tikzpicture}
        \begin{axis}[
            height=\smallplotheight,
            width=0.96*\smallplotwidth,
            legend pos= north east,
            legend style={font=\scriptsize, at={(1.02, 1.0)}, anchor=north west, legend columns=1, cells={anchor=west}},
            label style={font=\scriptsize},
            y label style={at={(-0.1,0.5)}},
            tick label style={font=\scriptsize},
            title style={align=left, font=\scriptsize},
            xmin=0.6,
            xmax=1,
            ymin=0.0,
            ymax=1,
            xlabel={normalized spectral efficiency, $s$},
            ylabel={$P(\text{nSE}>s)$},
            grid=both,
            yshift=-3.5cm,
        ]
            \addplot[gmmpgdulfromy, TUMBeamerBlue]
                table[x=GMM_PGD_UL_from_y,y=oneminy_label,col sep=comma] {gmm_feedback/P4/UL_all_se_rel_opt_cdf_gap40_nc64_SNR15dB_ntrain20000.csv};
                \addlegendentry{kGMM, $\mby$, $n_p$=$4$};
            \addplot[gmmpgdulfromy, TUMBlue, dashed]
                table[x=GMM_PGD_UL_from_y,y=oneminy_label,col sep=comma] {gmm_feedback_variableB/B6toep/P4/UL_all_se_rel_opt_cdf_gap40_nc64_SNR15dB_ntrain20000.csv};
                \addlegendentry{tGMM, $\mby$, $n_p$=$4$};
            \addplot[gmmpgdulfromy,  TUMBlue, dotted]
                table[x=GMM_PGD_UL_from_y,y=oneminy_label,col sep=comma] {gmm_feedback_variableB/B6circ/P4/UL_all_se_rel_opt_cdf_gap40_nc64_SNR15dB_ntrain20000.csv};
                \addlegendentry{cGMM, $\mby$, $n_p$=$4$};
            \addplot[lloydpgdulgmmulall,dash dot]
                table[x=Lloyd_PGD_UL_gmmULfull,y=oneminy_label,col sep=comma] {gmm_feedback/P4/UL_all_se_rel_opt_cdf_gap40_nc64_SNR15dB_ntrain20000.csv};
                \addlegendentry{\pltlloydpgdulgmmulall, $n_p$=$4$};
                
            \addplot[gmmpgdulfromy, TUMBeamerGreen]
                table[x=GMM_PGD_UL_from_y,y=oneminy_label,col sep=comma] {gmm_feedback/P8/UL_all_se_rel_opt_cdf_gap40_nc64_SNR15dB_ntrain20000.csv};
                \addlegendentry{kGMM, $\mby$, $n_p$=$8$};
            \addplot[gmmpgdulfromy, ourdarkgreen, dashed]
                table[x=GMM_PGD_UL_from_y,y=oneminy_label,col sep=comma] {gmm_feedback_variableB/B6toep/P8/UL_all_se_rel_opt_cdf_gap40_nc64_SNR15dB_ntrain20000.csv};
                \addlegendentry{tGMM, $\mby$, $n_p$=$8$};
            \addplot[gmmpgdulfromy,  ourdarkgreen, dotted]
                table[x=GMM_PGD_UL_from_y,y=oneminy_label,col sep=comma] {gmm_feedback_variableB/B6circ/P8/UL_all_se_rel_opt_cdf_gap40_nc64_SNR15dB_ntrain20000.csv};
                \addlegendentry{cGMM, $\mby$, $n_p$=$8$};
            \addplot[lloydpgdulgmmulall,gray,dash dot]
                table[x=Lloyd_PGD_UL_gmmULfull,y=oneminy_label,col sep=comma] {gmm_feedback/P8/UL_all_se_rel_opt_cdf_gap40_nc64_SNR15dB_ntrain20000.csv};
                \addlegendentry{\pltlloydpgdulgmmulall,$n_p$=$8$};
        \end{axis}
    \end{tikzpicture}

%% file: fig_mumimo_eccdf_iter.tex
    \begin{tikzpicture}
        \begin{axis}[
            height=\normalplotheight,
            width=\normalplotwidth,
            legend pos= north east,
            legend style={font=\scriptsize, at={(0.46,1.075)}, anchor=south, legend columns=3},
            label style={font=\scriptsize},
            tick label style={font=\scriptsize},
            y label style={at={(-0.075,0.5)}},
            title style={align=left, font=\scriptsize},
            xmin=2,
            xmax=16,
            ymin=0,
            ymax=1,
            xlabel={Sum-Rate, $s$, [bps$/\SI{}{\hertz}$]},
            ylabel={$P(\text{SR}>s)$},
            grid=both,
        ]
            \addplot[MUswmmsegmmpgdulfromy, solid, TUMBeamerBlue]
                table[x=SWMMSE_observation,y=oneminy_label,col sep=comma] {gmm_feedback_MU/WMMSE_new/P8d4/all_sumrate_cdf_gap40_nc64_SNR5dB_ntrain20000.csv};
                \addlegendentry{kGMM samples, $\mby$};
            \addplot[MUswmmsegmmpgdulfromy, TUMBlue, dashed]
                table[x=SWMMSE_observation,y=oneminy_label,col sep=comma] {gmm_feedback_MU_variableB/WMMSE/P8d4d1/merged/all_sumrate_cdf_gap40_nc64_SNR5dB_ntrain20000.csv};
                \addlegendentry{mtGMM samples, $\mby$};
            \addplot[MUswmmsegmmpgdulfromy, TUMBlue, dotted]
                table[x=SWMMSE_observation,y=oneminy_label,col sep=comma] {gmm_feedback_MU_variableB/WMMSE/P8d4d1/pruned/all_sumrate_cdf_gap40_nc64_SNR5dB_ntrain20000.csv};
                \addlegendentry{ptGMM samples, $\mby$};


            \addplot[MUgmmpgdulfromy]
                table[x=WMMSE_gmm_from_y,y=oneminy_label,col sep=comma] {gmm_feedback_MU/WMMSE_new/P8d1/all_sumrate_cdf_gap40_nc64_SNR5dB_ntrain20000.csv};
                \addlegendentry{kGMM, $\mby$};
            \addplot[MUgmmpgdulfromy, ourdarkgreen, dashed]
                table[x=WMMSE_gmm_from_y,y=oneminy_label,col sep=comma] {gmm_feedback_MU_variableB/WMMSE/P8d4d1/merged/all_sumrate_cdf_gap40_nc64_SNR5dB_ntrain20000.csv};
                \addlegendentry{mtGMM, $\mby$};
            \addplot[MUgmmpgdulfromy, ourdarkgreen, dotted]
                table[x=WMMSE_gmm_from_y,y=oneminy_label,col sep=comma] {gmm_feedback_MU_variableB/WMMSE/P8d4d1/pruned/all_sumrate_cdf_gap40_nc64_SNR5dB_ntrain20000.csv};
                \addlegendentry{ptGMM, $\mby$};


            \addplot[MUlloydpgdulgmmulall, dash dot]
                table[x=WMMSE_lloyd_from_hgmm,y=oneminy_label,col sep=comma] {gmm_feedback_MU/WMMSE_new/P8d1/all_sumrate_cdf_gap40_nc64_SNR5dB_ntrain20000.csv};
                \addlegendentry{\pltMUlloydpgdulgmmulall};
            \addplot[MUrandomgmmulall, dash dot]
                table[x=WMMSE_randomcb_from_hgmm,y=oneminy_label,col sep=comma] {gmm_feedback_MU/WMMSE_new/P8d1/all_sumrate_cdf_gap40_nc64_SNR5dB_ntrain20000.csv};
                \addlegendentry{\pltMUrandomgmmulall};
        \end{axis}
    \end{tikzpicture}

%% file: fig_mumimo_oversnr.tex
    \begin{tikzpicture}
        \begin{axis}[
            height=0.825*\largeplotheight,
            width=\largeplotwidth,
            legend pos= north east,
            legend style={font=\scriptsize, at={(0.475,1.025)}, anchor=south, legend columns=3},
            label style={font=\scriptsize},
            tick label style={font=\scriptsize},
            y label style={at={(-0.06,0.5)}},
            title style={align=left, font=\scriptsize},
            xmin=0,
            xmax=25,
            ymin=2.0,
            ymax=26,
            ytick={2,5,10,15,20,25},
            xlabel={SNR [$\SI{}{dB}$]},
            ylabel={Sum-Rate [bps$/\SI{}{\hertz}$]},
            grid=both,
        ]
            \addplot[MUswmmsegmmpgdulfromy, TUMBeamerBlue, solid,mark=square,mark size=2pt]
                table[x=snr,y=SWMMSE_observation] {gmm_feedback_MU/WMMSE_new/P8dBestscov/sumrate_over_snr_dBEST.txt};
                \addlegendentry{kGMM samples, $\mby$};
                
            \addplot[MUswmmsegmmpgdulfromy, TUMBlue, dashed, mark=square,mark size=2pt]
                table[x=snr,y=SWMMSE_observation,col sep=comma] {gmm_feedback_MU_variableB/WMMSE/P8d4d1/merged/sumrate_over_snr.csv};
                \addlegendentry{mtGMM samples, $\mby$};
            \addplot[MUswmmsegmmpgdulfromy, TUMBlue, dotted, mark=square,mark size=2pt]
                table[x=snr,y=SWMMSE_observation, col sep=comma] {gmm_feedback_MU_variableB/WMMSE/P8d4d1/pruned/sumrate_over_snr.csv};
                \addlegendentry{ptGMM samples, $\mby$};
                
            
            \addplot[MUgmmpgdulfromy,mark=o,mark size=2pt]
                table[x=snr,y=WMMSE_eigsp_gmm_fromy] {gmm_feedback_MU/WMMSE_new/P8dBestscov/sumrate_over_snr_dBEST.txt};
                \addlegendentry{kGMM, $\mby$};

            \addplot[MUgmmpgdulfromy, ourdarkgreen, dashed, mark=o,mark size=2pt]
                table[x=snr,y=WMMSE_eigsp_gmm_fromy,col sep=comma] {gmm_feedback_MU_variableB/WMMSE/P8d4d1/merged/sumrate_over_snr.csv};
                \addlegendentry{mtGMM, $\mby$};
            \addplot[MUgmmpgdulfromy,  ourdarkgreen, dotted, mark=o,mark size=2pt]
                table[x=snr,y=WMMSE_eigsp_gmm_fromy,col sep=comma] {gmm_feedback_MU_variableB/WMMSE/P8d4d1/pruned/sumrate_over_snr.csv};
                \addlegendentry{ptGMM, $\mby$};

            \addplot[MUlloydpgdulgmmulall,dash dot,mark=x,mark size=2pt]
                table[x=snr,y=WMMSE_eigsp_lloyd_fromhgmm] {gmm_feedback_MU/WMMSE_new/P8dBestscov/sumrate_over_snr_dBEST.txt};
                \addlegendentry{\pltMUlloydpgdulgmmulall};
            \addplot[MUlloydpgdulomp,dash dot,mark=diamond,mark size=2pt]
                table[x=snr,y=WMMSE_eigsp_lloyd_fromhomp] {gmm_feedback_MU/WMMSE_new/P8dBestscov/sumrate_over_snr_dBEST.txt};
                \addlegendentry{\pltMUlloydpgdulomp};
            \addplot[MUlloydpgdulscov,dash dot,mark=triangle,mark size=2pt]
                table[x=snr,y=WMMSE_eigsp_lloyd_fromhscov] {gmm_feedback_MU/WMMSE_new/P8dBestscov/sumrate_over_snr_dBEST.txt};
                \addlegendentry{\pltMUlloydpgdulscov};
            \addplot[MUrandomgmmulall,dash dot, mark=pentagon,mark size=2pt]
                table[x=snr,y=WMMSE_eigsp_randomcb_fromhgmm] {gmm_feedback_MU/WMMSE_new/P8dBestscov/sumrate_over_snr_dBEST.txt};
                \addlegendentry{\pltMUrandomgmmulall}
            \draw [TUMBeamerBlue,-{Stealth}, line width=\lineWidth] (14.3,20)--(12.3,20);
            \node at (13.3,21) {\scriptsize Samples};
            \draw [TUMBeamerGreen,-{Stealth}, line width=\lineWidth] (15.7,20)--(17.7,20);
            \node at (17.4,21) {\scriptsize Directions};
        \end{axis}
    \end{tikzpicture}